# Improper ferroelectricity in ultrathin hexagonal ferrites film

Xin Li,[1] Yu Yun,[1] Xiaoshan Xu[1,2]*

[1]Department of Physics and Astronomy, University of Nebraska, Lincoln, Nebraska 68588, USA
[2]Nebraska Center for Materials and Nanoscience, University of Nebraska, Lincoln, Nebraska 68588, USA
*Corresponding author: Xiaoshan Xu (X.X.)

**Abstract:**

The suppression of ferroelectricity in ultrathin films of improper ferroelectric hexagonal ferrites or manganites has been attributed to the effect of interfacial clamping, however, the quantitative understanding and related phenomenological model are still lacking. In this work, we report the paraelectric-to-ferroelectric phase transition of epitaxial h-$ScFeO_3$ films with different thickness through in-situ reflection high-energy electron diffraction (RHEED). Based on the interfacial clamping model and the Landau theory, we show that the thickness-dependence of the ferroelectric Curie temperature can be understood in terms of the characteristic length of interfacial clamping layer and the bulk Curie temperature. Furthermore, we found that the critical thickness of improper ferroelectricity is proportional to the characteristic length of interfacial clamping layer. These results reveal the essential role of mechanical clamping from interface on the improper ferroelectricity of hexagonal ferrites or manganites, and could serve as the guidance to achieve robust improper ferroelectricity in ultrathin films.

## Introduction

The two-dimensional (2D) ferroelectricity has attracted intensive research interests recently due to novel mechanisms for stabilizing polar order as well as their great potentials to device scalability.[1-3]. In principle, epitaxial films of bulk ferroelectrics can also be scaled down to several unit cells. However, for proper ferroelectrics, such as BaTiO$_3$ (BTO), PbTiO$_3$ (PTO) [4-8] of perovskite structures, if the depolarization field is not fully screened, there may exist critical thickness below which ferroelectricity is quenched. The critical-thickness problem can be absent in improper ferroelectrics. For example, in improper ferroelectric hexagonal manganites (h-$R$MnO$_3$, $R$=Sc,Y, Ho-Lu ) or ferrites (h-$R$FeO$_3$), ferroelectric order originates from the linear coupling of polarization with non-polar structural distortion. This mechanism enables ferroelectricity in the ultrathin limit[9-11], which is comparable to the range of thickness for 2D ferroelectricity.

On the other hand, recent studies revealed an interfacial clamping layer in hexagonal manganite and ferrite thin films [9,10], which may affect the ferroelectricity significantly in the ultrathin limit. The transition temperature from paraelectric to ferroelectric phase was observed to decrease with reducing thickness and the interfacial clamping effect suppresses polarization within the first 2 unit cells (uc) in h-RMnO$_3$ film[9]. Despite these observations, how the interfacial layer determines the temperature and thickness dependence of ferroelectricity has not been studied systematically, the ultrathin films of h-RFeO$_3$ could provide the chance to further clarify the general role of interfacial clamping layer.

Improper ferroelectric h-$R$FeO$_3$ is formed by triangle lattice of FeO$_5$ bipyramids sandwiched by rare earth layers[11,12]. The triangle lattice of FeO$_5$ rotates 60 deg for alternative half unit cell, as shown in Fig.1 (a). The collective displacement of FeO$_5$ bipyramids corresponds to the K$_3$ mode distortion. The K$_3$ mode causes the imbalanced displacements of ions along the $c$ axis ($\Gamma_2^-$ mode), leading to the ferroelectricity in h-$R$FeO$_3$. The ferroelectricity is called improper since the primary order parameters are $(Q, \phi_Q)$, in which $Q$ is the magnitude of in-plane displacement and $\phi_Q$ is the rotation angle of apical oxygen relative to the coordinates of FeO$_5$ bipyramids. The induced spontaneous polarization is proportional to $Q^3 \cos(3\phi_Q)$ in h-$R$FeO$_3$.

As shown in Fig.1(b), below $T_c$, the energy landscape of ferroelectric h-$R$FeO$_3$ is Mexican-hat shape with six-fold symmetry based on the Landau theory. For the ground states, the order parameter $\phi_Q$ can only take discrete values $\frac{n\pi}{3}$ where $n$ is integer. When the temperature increases, the magnitude of $Q$ for the ground states decrease gradually, which also leads to the suppression of polarization[13]. Above $T$c, the ferroelectric phase transforms to the paraelectric phase ($Q = 0$), as shown by in Fig.1(c) based on Landau theory[14]. In additional to temperature-dependent phase transition, decreasing thickness is also expected to trigger the ferroelectric-to-paraelectric transition considering the interfacial clamping, as depicted in Fig.1(d). While the thermally driven transition is intrinsic to the temperature dependence of free energy, the thickness-driven transition is expected to be an extrinsic effect due to the elastic disruption of polar order at the film/substrate interface.

Here we focus on h-ScFeO$_3$ films, in which the smaller radius of Sc ion leads to stronger K$_3$ distortion with $Q \approx 1$ angstrom[15,16]. We show that $T_c$ of the h-ScFeO$_3$ films, inferred from characteristic diffraction streaks of the in-situ RHEED patterns, decreases when the film thickness reduces. The thickness dependence of $T_c$ and a critical thickness of ferroelectricity ($\zeta_{FE}$) can be well modelled with the Landau theory using the characteristic length of the interfacial clamping layer ($\zeta_0$) and the bulk Curie temperature ($T_S$) as two key parameters. These results elucidate that, in hexagonal ferrites and manganites, the interfacial-clamping-originated critical thickness is expected to affect scalability when it is comparable to the thickness of 2D ferroelectric layers.

**Results and Discussion:**

To reveal the origin of thickness-dependent scaling of $T_c$, the epitaxial h-ScFeO$_3$ films with different thickness were grown using pulsed laser deposition (PLD), followed by annealing at high temperature. The detailed growth conditions can be found in the previous work [15]. As shown in Fig. 2(a), the 2θ scans of x-ray diffractions indicate that h-ScFeO$_3$ films along the (0001) direction were formed on both Al$_2$O$_3$ (0001) and SrTiO$_3$ (STO)(111) substrates without impurity phase. Moreover, the in-plane epitaxy relationship can be inferred by comparing the RHEED patterns of film with substrate in Fig. 2(b). Specifically, the in-plane [100] direction of h-ScFeO$_3$ film is parallel to $[11\bar{2}0]$ of Al$_2$O$_3$ and $[\bar{2}11]$ of STO, respectively.

In h-$R$FeO$_3$, paraelectric-to-ferroelectric phase transition is accompanied by the tripling of the in-plane unit cell for the ferroelectric phase (P6$_3$cm) compared with the paraelectric phase (P6$_3$/mmc), as indicated in Fig. 1(d). Since the separation of the RHEED streaks is inversely proportional to the in-plane lattice constant[17-19], RHEED patterns with the electron beam along the h-ScFeO$_3$ [100] direction provides a measurement for tripled unit cell of ferroelectric phase with P6$_3$cm symmetry. Specifically, the RHEED pattern of the paraelectric phase consists of the (0,0) and ($\pm n$,0) diffraction streaks. The tripled in-plane unit cell of the ferroelectric phase leads to the formation of weaker diffraction streaks at the $\pm(3n+1)/3$ and $\pm(3n+2)/3$ positions along h-ScFeO$_3$ [001] direction, as shown in Fig. 2(b).

Fig. 3(a) to (c) display two-dimensional RHEED patterns of 2, 5 and 17 unit cell (uc) h-ScFeO$_3$/Al$_2$O$_3$ films respectively, which were captured at room temperature, below which are the normalized line profiles of RHEED intensity after integrated along vertical direction, with background subtracted. For the 17 uc film, the intensity of the weak streaks gradually increases when the temperature decreases, indicating the existence of thermal driven paraelectric-to-ferroelectric phase transition. When the thickness is 2 uc, as shown in Fig. 3(a), no weak streaks could be identified down to the room temperature. The missing weak streaks suggests that the mechanical boundary conditions at the interface, or interfacial clamping, modifies the energy-minimum state and fully suppress the K$_3$ distortion in the interfacial layers.

To trace the change of $T_c$ with respective to the film thickness, the temperature dependence of RHEED intensity of the weak streaks was analyzed quantitatively. Fig. 4 shows the intensity of the weak streaks, normalized using the (1,0) and (-1,0) streaks, as a function of temperature for various film thickness for h-ScFeO$_3$ films grown on both Al$_2$O$_3$ and STO substrates. As shown in Fig. 4(a), there is no obvious transition for the intensity of the weak streak for the 2 uc film, so the ferroelectric $T_c$ may be lower than the room temperature. All the other films, with thickness ranging from 4 to 41 uc, exhibit a transition of the weak streak intensity above room temperature. More importantly, $T_c$ increases when the thickness increases.

To interpret the key factors determining the thickness-dependent scaling of $T_c$ in ultrathin h-$R$FeO$_3$ film and the potential critical thickness, we treat the h-$R$FeO$_3$ near the interface as interfacial clamping layer, where the magnitude of structural distortion ($Q$) is suppressed completely at the beginning, and increases gradually when rare earth layer is away from the interface[10,11]. Considering the free energy of the structural distortion and the elastic energy at the same time within the framework of Landau theory[14,20] (see details in supplementary), the structural distortion with thickness and temperature can be written analytically as:

$$Q(z,T) = Q_\infty(T) \frac{1-\exp\left(-\frac{z}{\zeta(T)}\right)}{1+\exp\left(-\frac{z}{\zeta(T)}\right)}, \text{with} \begin{cases} Q_\infty(T) = \sqrt{\frac{-a_0}{b}}\sqrt{1-\frac{T}{T_S}} = Q_0\sqrt{1-\frac{T}{T_S}} \\ \zeta(T) = \sqrt{\frac{k}{-a_0}}\sqrt{\frac{T_S}{T_S-T}} = \zeta_0\sqrt{\frac{T_S}{T_S-T}} \end{cases} \quad (1)$$

in which $a_0<0$ and $b>0$ are coefficients of Landau theory, $k$ corresponds to stiffness coefficient, $T_s$ is the Curie temperature of bulk state, and $\zeta_0$ is the characteristic length of interfacial clamping layer at $T = 0$ K. By integrating the energy density with the thickness of ferroelectric layer ($t_F$), the total energy can be expressed as:

$$F(t_F,T) = -\frac{a_0}{2T_S} C1(t_F,T) \left\{ \left[T - \left(T_S + \frac{2kT_S}{a_0}\frac{C2(t_F,T)}{C1(t_F,T)}\right)\right] Q_\infty(T)^2 + \frac{T_S}{2*Q_0^2}\frac{C3(t_F,T)}{C1(t_F,T)} Q_\infty(T)^4 \right\} \quad (2)$$

with

$$\begin{cases} C1(t_F,T) = \int_0^{t_F} \left(\frac{1-\exp\left(-\frac{z}{\zeta(T)}\right)}{1+\exp\left(-\frac{z}{\zeta(T)}\right)}\right)^2 dz \\ C2(t_F,T) = \int_0^{t_F} \frac{\left(\frac{2}{\zeta}*\exp\left(-\frac{z}{\zeta(T)}\right)\right)^2}{\left(1+\exp\left(-\frac{z}{\zeta(T)}\right)\right)^4} dz \\ C3(t_F,T) = \int_0^{t_F} \left(\frac{1-\exp\left(-\frac{z}{\zeta(T)}\right)}{1+\exp\left(-\frac{z}{\zeta(T)}\right)}\right)^4 dz \end{cases} \quad (3)$$

When the coefficient before $Q_\infty(T)^2$ term is less than zero, minimizing $F(t_F,T)$ results in finite $Q_\infty(T)$ corresponding to the ferroelectric order. At $T=T_c$, this coefficient is zero, i.e.,

$$\alpha(t_F,T) = T - T_S\left[1 - 2\zeta_0^2 \frac{C2(t_F,T)}{C1(t_F,T)}\right] = 0 \quad (4)$$

(see details in supplementary).

The phase diagram of $\alpha(t_F, T)$ for h-ScFeO$_3$ films is plotted in Fig. 5(a), with $\zeta_0$ and $T_S$ inferred in Fig.5(b). Since $\alpha(t_F, T) < 0$ ($> 0$) corresponds to the ferroelectric (paraelectric) state of the h-$R$FeO$_3$ films, the curve of $\alpha(t_F, T) = 0$ indicates that the critical thickness of ferroelectricity increases with temperature, which is consistent with the observation of temperature-dependent corrugation of rare earth layers in h-YMnO$_3$/YSZ film [10]. Moreover, as implied by Eq. (4), the dependence of $T_c$ on $t_F$ is influenced by both $T_S$ and $\zeta_0$, which are two independent parameters that can be extracted from experimental observations, such as in-situ RHEED patterns in this work. As shown in Fig. 5(b), the experimental $T_c$ of both h-ScFeO$_3$ and h-YMnO$_3$ films decreases with smaller thickness, and the trend is slower in the h-ScFeO$_3$ films. By fitting the experimental data with Eq. (4), we find for h-YMnO$_3$/YSZ film, that $T_s = 814 \pm 183$ K and $\zeta_0 = 0.93 \pm 0.44$ uc. For the h-ScFeO$_3$ films, we find $T_s = 681 \pm 28$ K and $\zeta_0 = 0.68 \pm 0.18$ u.c. Therefore, the slower suppression of $T_c$ with thickness in h-ScFeO$_3$ films can be attributed to smaller $\zeta_0$.

It should be noted that the phenomenological model here does not consider the transition of single domain to multi-domain states with thickness, since the elastic energy is much larger than depolarization energy [13], as well as possible variation of stiffness constant at the interface, which may contribute to the minor discrepancy. Moreover, based on Eq. (4), the ferroelectric critical thickness ($\zeta_{FE}$) can be defined as the interception of $\alpha(t_F, T) = 0$ with $T = 0$ K, as shown in Fig.5 (c), and the ratio of $\zeta_{FE}/\zeta_0 = 2.25$ is a fixed value, which is independent of materials (see details in supplementary Section 3). This ratio indicates that the absence of ferroelectric critical thickness ($\zeta_{FE} = 0$) or unsuppressed corrugation of initial rare-earth layer in h-RFeO$_3$ or h-RMnO$_3$ films can only exist when there is no interfacial clamping layer ($\zeta_0 = 0$) or when the films become freestanding. Moreover, the improper ferroelectricity under ultrathin limit could be artificially controlled by choosing the characteristic length of interfacial clamping layer through different interface.

**Conclusion**

In summary, through quantitative study of in-situ RHEED patterns during paraelectric-to-ferroelectric phase transition, the thickness-dependent scaling of ferroelectric $T_c$ in epitaxial h-ScFeO$_3$ films was revealed for the first time. Based on interfacial clamping layer, a phenomenological model from Landau theory was introduced to interpret this scaling behavior and reveal the correlation between ferroelectric critical thickness and characteristic length of interfacial clamping layer in ultrathin h-RFeO$_3$ or h-RMnO$_3$ film. These results serve as the guidance to achieve robust improper ferroelectricity in h-RFeO$_3$ through interfacial engineering for future study.

**Experimental section**

The h-ScFeO$_3$/Al$_2$O$_3$ (0001) and h-ScFeO$_3$/SrTiO$_3$ (111) epitaxial thin films were grown using pulsed laser polarization (PLD) with base pressure lower than $3\times10^{-7}$ mTorr, a repetition rate of 2 Hz. Before the deposition, the substrates were pre-annealed at 700 ºC for 1 hour. During the growth, the substrate temperature was kept at 700 ºC - 920 ºC, specifically, the films were grown at low temperature first (~ 700 ºC) then annealed at high temperature (~ 920 ºC). The growth oxygen pressure is 10 mTorr. The in-situ RHEED was used to monitor ferroelectric to paraelectric phase transition of h-ScFeO$_3$ film after the growth, the incident angle of electron beam is kept fixed during the process of reducing temperature, and the raw images of RHEED are analyzed based on python program. The crystal structure and the thickness of the epitaxial h-ScFeO$_3$ films were measured by XRD (Rigaku SmartLab Diffractometer).

**Acknowledgements**

Funding: This work was primarily supported by the National Science Foundation (NSF), Division of Materials Research (DMR) under Grant No. DMR-1454618 and by the Nebraska Center for Energy Sciences Research. The research was performed in part in the Nebraska Nanoscale Facility: National Nanotechnology Coordinated Infrastructure and the Nebraska Center for Materials and Nanoscience, which are supported by the NSF under Grant No. ECCS- 2025298, and the Nebraska Research Initiative.

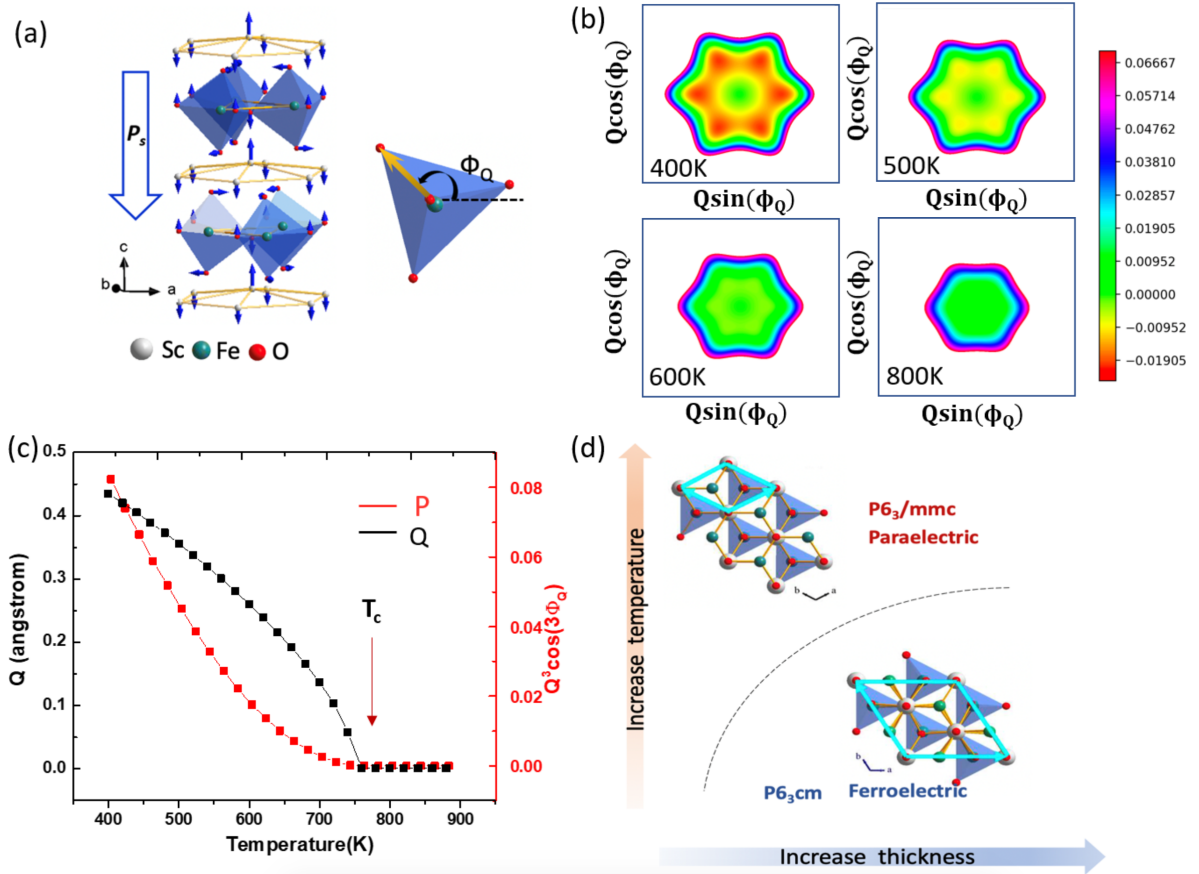

Fig.1 (a) Atomic structure of h-$R$FeO$_3$ and the rotation angle of apical oxygen atoms ($\phi_Q$). (b) Representative energy landscapes of h-$R$FeO$_3$ at different temperature, assuming $T_c$ as 750 K. (c) The schematic diagram for the temperature dependence of the order parameter $Q$ and the polarization $P=Q^3\cos(3\phi_Q)$ for bulk-state h-$R$FeO$_3$, based on the Landau theory. (d) Schematic diagrams for the temperature and thickness driven phase transition in h-RFeO$_3$ films and related atomic structures in a-b plane.

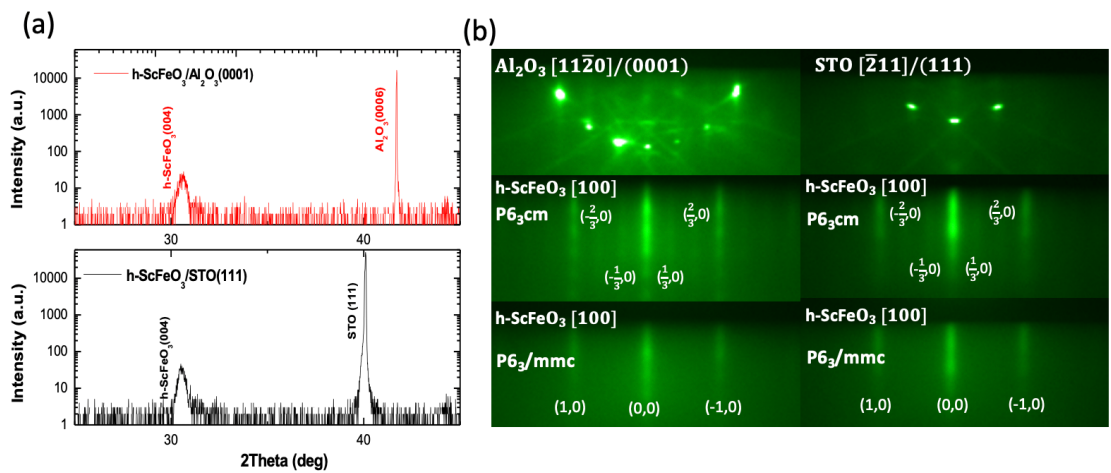

Fig.2 (a) 2θ scan for h-ScFeO$_3$/Al$_2$O$_3$ and (b) h-ScFeO$_3$/STO films. (b) RHEED patterns of ferroelectric phase (P6$_3$cm symmetry) and paraelectric phase (P6$_3$/mmc symmetry) and in-plane epitaxy relationships for h-ScFeO$_3$/Al$_2$O$_3$ and h-ScFeO$_3$/STO films.

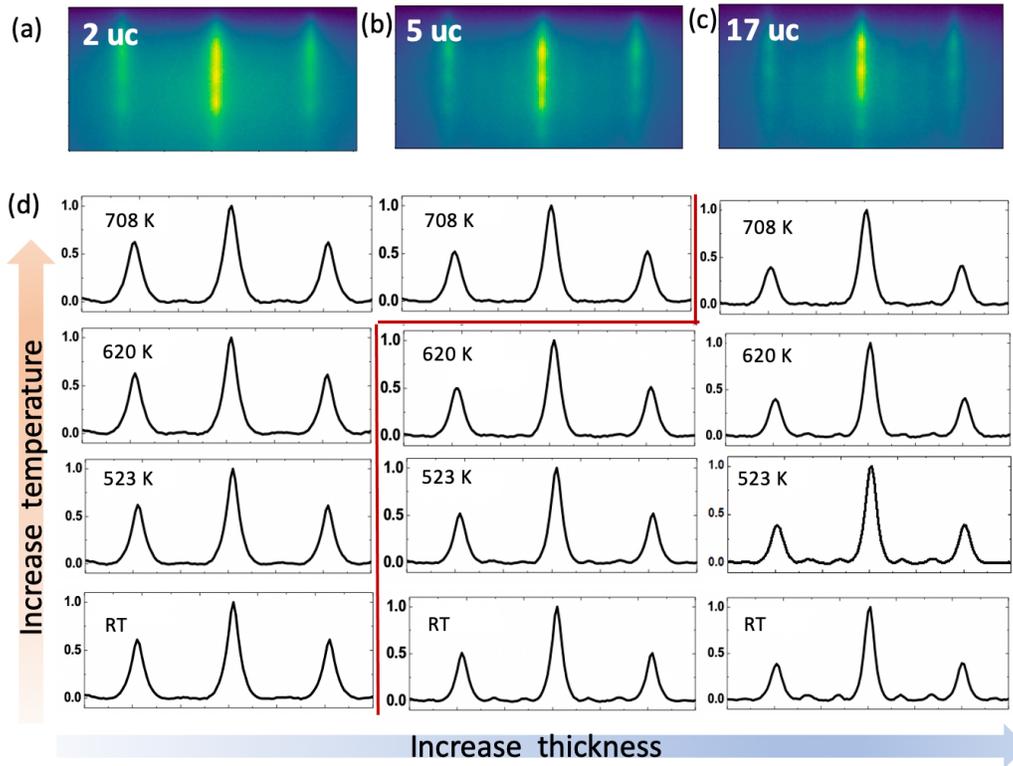

Fig.3 (a) to (c) RHEED images of h-ScFeO$_3$/Al$_2$O$_3$ films with thickness of 2, 5, 17 uc at room temperature. (d) Normalized profiles of RHEED intensity with different thickness and temperature for h-ScFeO$_3$/Al$_2$O$_3$ films.

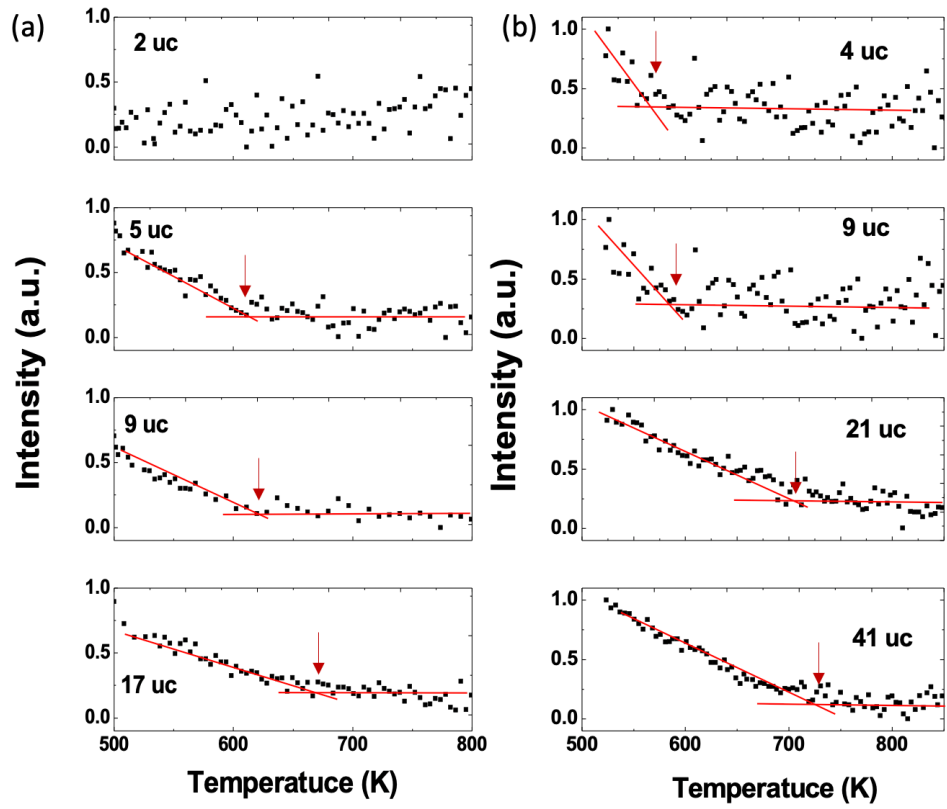

Fig.4 Normalized RHEED intensity of (1/3, 0) streak with temperature for (a) h-ScFeO$_3$/Al$_2$O$_3$ and (b) h-ScFeO$_3$/STO films with different thickness.

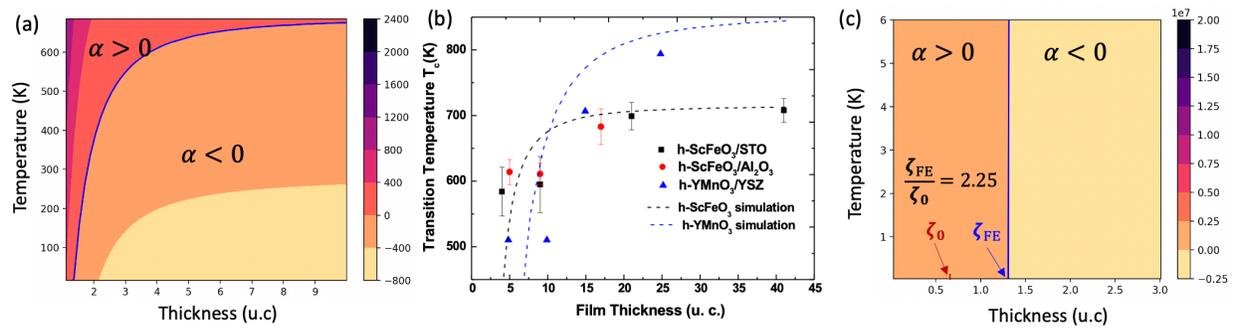

Fig.5 (a) Calculated phase diagram of coefficient α in Eq. (4) with temperature and film thickness as independent variables, the blue solid blue line corresponds to α = 0. (b) Measured thickness-dependent $T$c of h-ScFeO$_3$/STO, h-ScFeO$_3$/Al$_2$O$_3$ and h-YMnO$_3$/YSZ films and related fitting based on Eq. (4), the data of h-ScFeO$_3$ comes from in-situ RHEED in this work and the data of h-YMnO$_3$ comes from Ref. [9]. (c) Phase diagram of α near $T$= 0 K and the comparison of $\zeta_0$ and $\zeta_{FE}$.